\documentclass[a4paper, amsfonts, amssymb, amsmath, reprint, showkeys, nofootinbib, twoside,notitlepage,onecolumn]{revtex4-2}

\def\prd{{Phys.~Rev.~D}}        
\def\prl{{Phys.~Rev.~Lett.}}    
\usepackage{amsmath,amstext}
\usepackage[T1]{fontenc}
\usepackage{amssymb}
\input{epsf}
\usepackage{graphicx}
\usepackage{ae,aecompl}

\usepackage{hyperref}
\usepackage{amsmath}
\usepackage{amssymb}
\usepackage{mathtools}
\usepackage{bm}
\usepackage{cleveref}
\usepackage{tensor}
\usepackage{braket}
\usepackage{enumitem}
\usepackage{mhchem}
\usepackage{cancel}
\usepackage{amsthm}
\usepackage{upgreek}
\usepackage{mathrsfs}

\usepackage{color}

\newcommand{\spc}{\quad \quad \quad}

\def\be{\begin{equation}}
\def\ee{\end{equation}}
\def\beq{\begin{eqnarray}}
\def\eeq{\end{eqnarray}}

\theoremstyle{definition}

\theoremstyle{theorem}
\newtheorem{theorem}{Theorem}

\theoremstyle{corollary}

\begin{document}
\title{Solitons and singularities in relativistic ultrastiff fluids}
\author{ L.~Gavassino}
\affiliation{
$^1$Department of Mathematics, Vanderbilt University, Nashville, TN, USA
}

\begin{abstract}
A mathematical duality exists between massless scalar fields and relativistic fluids governed by an ultrastiff equation of state, in which the pressure equals the mass-energy density, and the sound speed equals $c$. This duality entails that certain solutions of the wave equation (a linear theory) can be mapped to solutions of ideal relativistic hydrodynamics (an inherently non-linear theory). Leveraging this correspondence, we explore some interesting properties of ultrastiff fluids. Specifically, we demonstrate that all non-linear sound waves in such media are solitons. Moreover, we prove that, in 
3+1 dimensions, compactly supported configurations of an ultrastiff fluid inevitably evolve, in finite time, toward a singular state where the fluid’s flow velocity exits the future lightcone. We also demonstrate that, prior to the formation of this singularity, first-order perturbation theory in the small parameter $c_s^{-2}-c^{-2}$ produces divergent results. As a consequence, a fluid whose speed of sound is, say, $0.995 \, c$ exhibits markedly different behavior near the singularity compared to a fluid with a speed of sound exactly equal to $c$.
\end{abstract}

\maketitle

\section{Introduction}

There is a growing consensus that the speed of sound ($c_s$) of neutron-star matter is quite high, potentially exceeding 0.6 \cite{Altiparmak:2022bke,Koehn:2024set} (with $c=1$), though the extent of this excess remains uncertain. This finding strengthens the widespread belief that nothing prevents the speed of sound of dense matter from approaching the speed of light in some regions of the QCD phase diagram \cite{Zeldovich:1961sbr,SonPionCondensate2000by,MooreSpeedofSound2024gmt}\footnote{A novel upper bound (namely $c_s^2 \leq 0.781$) has been recently introduced in \cite{Hippert:2024hum}. However, that is not a statement about the \textit{non-existence} of fluids with larger speed of sound. It just provides a dynamical characterization of such fluids. In particular: Fluids with $c_s^2 > 0.781$ cannot be modeled using ordinary Israel-Stewart theory \cite{Israel_Stewart_1979,Hishcock1983}, or any kind of viscoelastic Maxwell-type model \cite{GavassinoGENERIC2022,GavassinoUniversalityI2023}. Currently, the only universally accepted \textit{existence} bound is $c_s^2\leq 1$, which can be derived directly from relativistic statistical mechanics \cite{GavassinoCausality2021,HellerBounds2022ejw,GavassinoBounds2023}.}. This is a rather intriguing possibility since, in a phase of this kind, sound waves would propagate with the same speed in all reference frames, potentially giving rise to unusual fluid-dynamical phenomena that are intrinsically relativistic. The goal of this article is to analyse some of these phenomena.

Most instances of fluid phases with luminal sound speed discussed in the literature \cite{Zeldovich:1961sbr,SonPionCondensate2000by,MooreSpeedofSound2024gmt} may be approximated as ``ultrastiff fluids'', namely fluids whose equation of state is approximately given by
\begin{equation}\label{Pe}
    P=\varepsilon  \spc (\text{so that } c_s^2=dP/d\varepsilon=1) \, ,
\end{equation}
where $P$ is the pressure, and $\varepsilon$ is the energy density. This type of matter is rather special because it saturates the dominant energy condition ($|P|\leq \varepsilon$ \cite[\S 4.3]{HawkingEllis1973uf}). In a sense, ultrastiff fluids may be viewed as the logical opposite of the ``dust'' (whose equation of state is $P=0$ \cite[\S 1.9]{carroll_2019}), which is an idealization for an ``ultrasoft'' material that does not resist compression. Here, we will model the ultrastiff phase as an ideal fluid, with stress-energy tensor\footnote{It was rigorously proven that a thermodynamically stable phase of matter with $c_s^2=1$ must have vanishing viscous corrections at all orders in the gradient expansion \cite{GavassinoBounds2023,HellerHydrohedron2023jtd}. This means that, if a fluid with the equation of state \eqref{Pe} exists, it can only be a perfectly inviscid fluid. In reality, a fluid with identically zero viscosity is unlikely to exist in a quantum world \cite{Kovtun:2004de}. For this reason, one should consider the equation of state \eqref{Pe} as an idealized limit (just like the dust).}
\begin{equation}\label{idealfluid}
    T^{\mu \nu}=(\varepsilon{+}P)u^\mu u^\nu +Pg^{\mu \nu} = P \, (2u^\mu u^\nu {+}g^{\mu \nu}) \, ,
\end{equation}
which is very convenient for the following ``mathematical'' reason: If the flow is irrotational (in a precise sense that we define in section \ref{dualz}), then it is possible to map the initial value problem for the conservation law $\partial_\mu T^{\mu \nu}=0$ into a dual initial value problem for a real scalar field $\Psi$ that obeys the linear wave equation:
\begin{equation}\label{waves}
    \partial_\mu \partial^\mu \Psi=0 \, .
\end{equation}
The existence of this duality is textbook material \cite[\S 3.7.4]{rezzolla_book}. However, to the best of our knowledge, its broad implications have not been fully appreciated till now. In fact, we recall that, due to their non-linear nature, the equations of relativistic fluid mechanics are usually impossible to solve analytically, except in highly symmetric scenarios. For (irrotational) ultrastiff matter, the non-linearity obstacle is immediately overcome, since we can just pick a solution to the wave equation (which is a linear equation), and ``reconstruct'' from it an exact solution of the ultrastiff fluid equations. We will use this trick to prove general statements concerning the unconventional behavior of these highly relativistic substances.

Throughout the article, we work in Minkowski spacetime, with signature $(-,+,+,+)$, and natural units: $c=k_B=1$.
\newpage


\section{Correspondence between ultrastiff fluids and scalar fields}\label{dualz}

In this section, we provide a simple derivation of the duality between massless scalar fields and ultrastiff fluids. Then, we discuss possible physical interpretations of these fluids, and of their scalar duals.  

\subsection{Quick derivation of the duality}

Given the stress-energy tensor \eqref{idealfluid}, let us define a new vector field $\xi^\mu{:=} \sqrt{P} \, u^\mu$, whose Minkowski length is $\xi^\lambda \xi_\lambda {=}-P$. Then, it is immediate to see that the stress-energy tensor is quadratic in this new field:
\begin{equation}\label{intermedio}
    T\indices{^\mu _\nu}=2\xi^\mu \xi_\nu{-}\xi^\lambda \xi_\lambda \, g\indices{^\mu _\nu} \, .
\end{equation}
The energy-momentum conservation law, $\partial_\mu T\indices{^\mu _\nu}=0$, can be projected parallelly and perpendicularly to $\xi^\mu$, giving
\begin{equation}\label{motionintermedio}
\begin{split}
\partial_\mu \xi^\mu ={}& 0 \, ,\\  
\xi^\mu \partial_{[\mu} \xi_{\nu]}={}& 0 \, . \\
\end{split}
\end{equation}
Let us focus on the implications of the second equation. Using the language of exterior calculus \cite{LoringTubook}, we can equivalently rewrite it in the notation $\iota_\xi d\xi=0$. Thus, using Cartan's magic formula, we obtain $\mathcal{L}_\xi d\xi=\iota_\xi d^2 \xi+d\iota_\xi d\xi=0$. This tells us that the ``vorticity tensor'' $\partial_{[\mu} \xi_{\nu]}$ is conserved along the flow generated by $\xi^\mu$ \cite{noto_rel}. It follows that, if $\partial_{[\mu} \xi_{\nu]}$ vanishes at $t=0$, it must vanish across the whole spacetime, in which case we can set $\xi_\mu =\partial_\mu \Psi$ for some real scalar field $\Psi$. This is what we mean by the flow being ``irrotational''. Replacing $\xi_\mu$ with $\partial_\mu \Psi$ in \eqref{intermedio} and \eqref{motionintermedio}, we finally obtain
\begin{equation}
\begin{split}\label{psiuzzuzu}
& T^{\mu \nu}=2\partial^\mu \Psi \, \partial^\nu \Psi -\partial^\lambda \Psi \partial_\lambda \Psi \, g^{\mu \nu} \, , \\
& \partial_\mu \partial^\mu \Psi=0 \, . \\
\end{split}
\end{equation}
These are the Noether stress-energy tensor and the Euler-Lagrange equation of a scalar field with Lagrangian density $\mathcal{L}=\partial_\mu \Psi \partial^\mu \Psi$ \cite{carroll_2019}. Hence, we have converted the ultrastiff fluid into a massless scalar field theory.

In the following, we will use the above duality in the other direction. Namely, we will compute analytical solutions $\Psi(x^\mu)$ of the wave equation, and will then reconstruct the associated fluid variables through the correspondence $\partial_\mu \Psi=\sqrt{P} \, u_\mu$, which results in the following ``dictionary formulas'':
\begin{equation}
\begin{split}
P ={}& - \partial^\lambda \Psi \partial_\lambda \Psi \, , \\ 
u^\mu ={}& \dfrac{\partial^\mu \Psi}{\sqrt{-\partial^\lambda \Psi \partial_\lambda \Psi}} \, . \\
\end{split}
\end{equation}
Note that, for the formal bridge to hold in the direction $\Psi\rightarrow \{P,u^\mu\}$, the vector $\partial^\mu \Psi$ must be timelike future-directed. In the following, we will always choose initial conditions (at $t=0$) that fulfill this requirement. However, it is not guaranteed that this same constraint will also hold at later times. This subtle issue will be studied in detail later.

\subsection{Thermodynamic interpretation}

General arguments suggest that a fluid in a quantum world cannot have vanishing shear viscosity, unless its entropy density vanishes \cite{Kovtun:2004de}. Hence, we believe that, if an ultrastiff fluid were to exist, it would most likely have zero temperature. Let $n$ be its baryon density, and $\mu$ be its chemical potential. Then, using the thermodynamic identities $d\varepsilon=\mu \, dn$ and $dP=n \, d\mu$, we obtain, from \eqref{Pe}, the differential equation $\mu \, dn = n \, d\mu$, whose solution is $\mu = 2K n$, for some integration constant $K$. Integrating in $n$, we obtain the following relations:
\begin{equation}\label{sgargiante}
    \varepsilon= Kn^2 =\dfrac{\mu^2}{4K} =P \, .
\end{equation}
The possible existence of zero-temperature matter with this equation of state has been studied by \citet{Zeldovich:1961sbr}. They found that, if the baryons interact through a massive vector field, then the equation of state has an asymptotic ultrastiff limit at ultrahigh densities. More recently, it was shown that the equation of state \eqref{sgargiante} naturally arises (as an asymptotic limit) in relativistic fluids with scalar condensates arising from a $O(N)$ broken symmetry \cite{MooreSpeedofSound2024gmt}.

Since $\sqrt{P}\propto n \propto \mu$, we can rewrite the equations $\partial_\mu \xi^\mu=0$ and $\xi_\mu =\partial_\mu \Psi$ in the following more intuitive form:
\begin{equation}
\begin{split}
& \partial_\mu (nu^\mu)=0 \, , \\
& \mu u_\mu =\partial_\mu (2\sqrt{K} \Psi)  \, . \\
\end{split}
\end{equation}
The first equation tells us that the baryon number is conserved, while the second tells us that the ``chemical momentum'' is the gradient of a scalar, which is just a restatement of the irrotationality assumption. The latter condition is known to occur in zero-temperature superfluids \cite{Carter_starting_point,cool1995,Son2001,Termo} (note that, at $T=0$, the normal component disappears), where $\Psi$ is proportional to the gradient of the phase of the order parameter. Then, the dual scalar-field theory is just the effective field theory of the massless Goldstone boson associated with the superfluid symmetry breaking. However, we remark that superfluidity is only a sufficient condition for irrotationality. It is not a necessary condition. Any zero-temperature ideal fluid whose initial state happens to be irrotational remains irrotational at all times.

\section{Superposition principle}\label{IIIyup}

The fluid equations, expressed in terms of $\{ P,u^\mu\}$, are non-linear. Therefore, if two fluids (or two sound waves) ``come across'' each other, we cannot just sum their respective solutions for $P$ and $u^\mu$. On the other hand, the equation for $\Psi$ is linear. Thus, we can add the corresponding scalar fields. Let us discuss the physical consequence of this fact.

\subsection{Mathematical statement}

Let $\Psi_1$ and $\Psi_2$ be two solutions of \eqref{waves}, and $\{P_1,u^\mu_1 \}$ and $\{P_2,u^\mu_2 \}$ the corresponding ultrastiff fluids. The superposition $\Psi=\Psi_1+\Psi_2$ is also a solution of \eqref{waves}. Let us compute the corresponding ``superimposed'' fluid variables $\{P,u^\mu \}$. Since $\partial_\mu \Psi=\partial_\mu \Psi_1+\partial_\mu \Psi_2$, we immediately obtain
\begin{equation}\label{ebuno}
    \sqrt{P} \, u^\mu = \sqrt{P}_1 \, u_1^\mu +\sqrt{P}_2 \, u_2^\mu \, .
\end{equation}
As can be seen, we are not adding up the pressures. Rather, we are adding up the baryon currents (recall that $\sqrt{P}\propto n$). Note that, if both $\partial^\mu \Psi_1$ and $\partial^\mu \Psi_2$ are timelike future-directed, then also their sum is timelike future-directed. In other words, if the two fields $\Psi_1$ and $\Psi_2$ admit an interpretation as fluids, then also their superposition is a valid fluid. Using \eqref{ebuno}, we can explicitly evaluate the superimposed pressure and flow velocity:
\begin{equation}\label{orso}
\begin{split}
P ={}& P_1+P_2 +2\gamma \sqrt{P_1 P_2} \, , \\
u^\mu={}& \dfrac{\sqrt{P}_1 \, u_1^\mu +\sqrt{P}_2 \, u_2^\mu}{\sqrt{P_1+P_2 +2\gamma \sqrt{P_1 P_2}}} \, , \\
\end{split}
\end{equation}
where $\gamma=-u^\mu_1 u_{2\mu}$ is the relative Lorentz factor of the two solutions. As can be seen, the total pressure is always larger than the sum of the superimposed pressures. This reflects the rigid nature of ultrastiff matter, which responds to an increase in density with much more pressure than an ideal gas (for which pressures are just added).

Let us now superimpose the stress-energy tensors directly. Replacing $\Psi$ with $\Psi_1+\Psi_2$ in \eqref{psiuzzuzu}, we obtain
\begin{equation}
T^{\mu \nu}= T_1^{\mu \nu}+T_2^{\mu \nu}+2\sqrt{P_1 P_2} \big[ u_1^\mu u_2^\nu{+}u_2^\mu u_1^\nu {+}\gamma g^{\mu \nu}\big] \, .
\end{equation}
Again, we see that the superimposed stress-energy tensor is not the sum of the two stress-energy tensors, since there is an additional bilinear term in $\sqrt{P}_1 \, u_1^\mu$ and $\sqrt{P}_2 \, u_2^\mu$.

\subsection{Collisions of two fluid drops}

Suppose that two ``drops'' of ultrastiff matter are traveling towards each other close to the speed of light. At some point, they collide. What happens afterwards? 

\begin{figure}
    \centering
\includegraphics[width=0.54\textwidth]{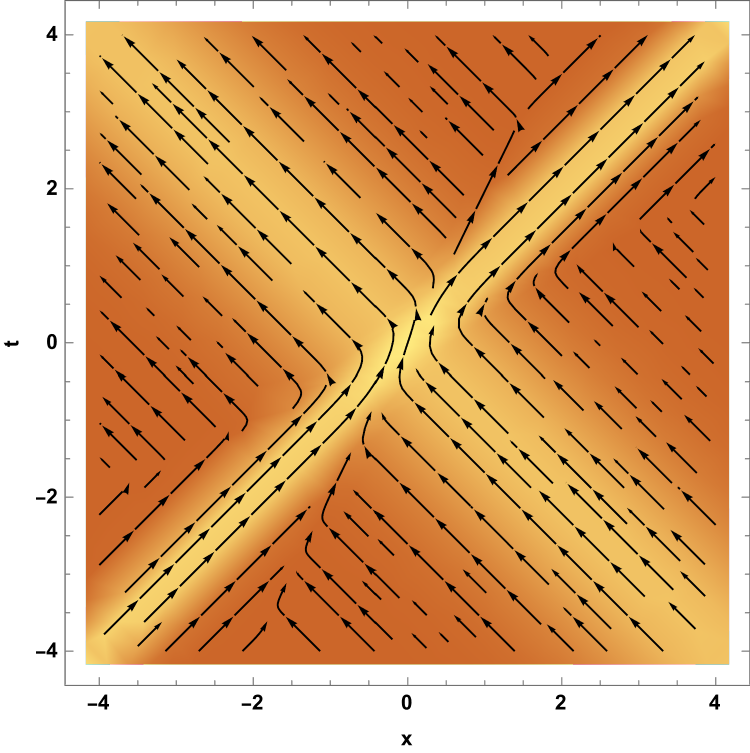}
    \caption{Minkowski diagram of a collision between two drops of ultrastiff matter in 1+1 dimensions. The colormap represents the density of particles in the given reference frame, namely $n u^0 \propto \sqrt{P} u^0=\partial^t \Psi$. If we focus on the colormap, it looks like the two drops are just going through each other. However, the flow worldlines (black arrows), i.e. the integral curves of $u^\mu$, show that the dynamics is more interesting. The left incoming drop is absorbed into the right incoming drop. During this process, a portion of the right drop is ``kicked out'' from the main stream, and is sent to the right. The result are two outgoing drops that are identical copies of the incoming ones. The scalar field dual to this process is $\Psi(t,x)=\frac{2}{3}\text{Erf}\big[3(x-t)\big]-\text{Erf}(x+t)$.}
    \label{fig:clumps}
\end{figure}

Let us formulate this problem as an ordinary scattering experiment \cite[Ch 3]{weinbergQFT_1995}. At $t=-\infty$, the two fluid drops are far apart, and their supports are disjoint. Thus, the scalar field $\Psi$ may be decomposed as the sum of two distinct parts, namely $\Psi(-\infty)=\Psi_1(-\infty)+\Psi_2(-\infty)$, representing the incoming states of the two drops. Since the wave equation is linear, and its solution is unique \cite[Theor 10.1.2]{Wald}, the scalar field $\Psi$ must be given by $\Psi=\Psi_1+\Psi_2$ at \textit{all} times, where $\Psi_1$ is the solution with early state $\Psi_1(-\infty)$, and $\Psi_2$ is the solution with early state $\Psi_2(-\infty)$. In other words, the scalar field $\Psi$ describing a collision between two drops of ultrastiff fluid is just the sum of the scalar fields $\Psi_1$ and $\Psi_2$ of the two freely traveling drops. This has a rather surprising implication: The drops effectively ``go past'' each other, and the outgoing states retain no memory of the collision. In fact, at $t=+\infty$, long after the collision has happened, $\Psi(+\infty)=\Psi_1(+\infty)+\Psi_2(+\infty)$
is still the sum of two distinct drops, whose shapes could be predicted by evolving $\Psi_1(-\infty)$ and $\Psi_2(-\infty)$ separately.
The above analysis may be summarised in a simple theorem.
\begin{theorem}\label{theo1}
Drops of irrotational ultrastiff matter have vanishing collision cross-section.   
\end{theorem}
Based on this result, it is tempting to conclude that ultrastiff matter is somehow ``transparent'', in the sense that the two drops do not ``feel'' each other when they meet. However, that is not the case. Recall that this type of matter is extremely rigid. When the two drops collide, they experience the highest pressure allowed by relativity. To see this, consider that, in the limit in which the two solutions $\Psi_1$ and $\Psi_2$ travel relatively to each other at the speed of light, the factor $\gamma$ in equation \eqref{orso} diverges. Therefore, at the time of the collision, we have that $P\rightarrow \infty$. Such an ultrahigh pressure results in a perfect pushback, which leads to the formation of two \textit{new} outgoing clumps, whose shapes happen to be identical to those of the incoming drops. Hence, what seems like transparent behavior is actually the strongest form of repulsion allowed in nature. An analytical example is provided in figure \ref{fig:clumps}.

\section{Non-linear solitary waves}

In this section, we discuss the propagation and interference of non-linear sound waves in ultrastiff fluids. Given the statement of Theorem \ref{theo1}, it is natural to expect that such waves should exhibit some soliton-like behavior, which is indeed what we aim to verify.

\subsection{General plane-wave solution}
\vspace{-0.3cm}

Let us consider the following class of solutions of the wave equation with planar symmetry:
\begin{equation}\label{waves}
\Psi(t,x)= -t A + \int_0^{x-t} f(L) dL -\int_0^{x+t} g(L) dL \, ,
\end{equation}
where $A>0$ is a constant, while $f$ and $g$ are two smooth functions. Since there is no dependence on $y$ and $z$, we have that $u^y \propto\partial^y \Psi =0$ and $u^z \propto\partial^z \Psi =0$. Thus, the solution \eqref{waves} describes a longitudinal wave, with
\begin{equation}
\begin{split}
\sqrt{P} u^t ={}& A+f(x{-}t)+g(x{+}t) \, , \\
\sqrt{P} u^x ={}& f(x{-}t)-g(x{+}t) \, . \\
\end{split}
\end{equation}
From this, we immediately obtain an expression for the pressure $P$ and the ordinary three-velocity $v=u^x/u^t$, namely
\begin{equation}
\begin{split}
P={}& A^2+2A\big[f(x{-}t)+g(x{+}t)]+4f(x{-}t)g(x{+}t)   \, , \\
v={}& \dfrac{f(x{-}t)-g(x{+}t)}{A+f(x{-}t)+g(x{+}t)} \, . \\
\end{split}
\end{equation}
These hydrodynamic solutions describe the exact propagation of planar-symmetric non-linear soundwaves in ultrastiff matter. The constant $A^2$ is the energy density of the unperturbed state. The function $f$ describes the right-traveling soundwaves, while the function $g$ describes the left-traveling soundwave. Note that, if $f$ and $g$ are sufficiently small compared to $A$, the pressure remains always positive, and the velocity remains bounded between $-1$ and $1$.

\vspace{-0.3cm}
\subsection{Solitonic behavior}
\vspace{-0.3cm}

\begin{figure}[h!]
    \centering
\includegraphics[width=0.55\textwidth]{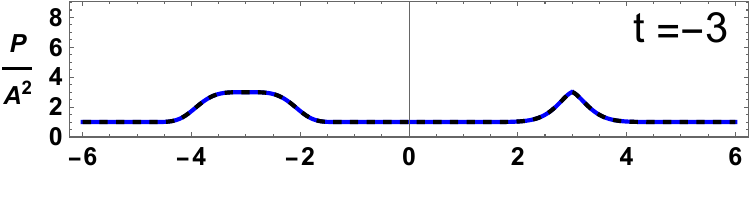}
\includegraphics[width=0.55\textwidth]{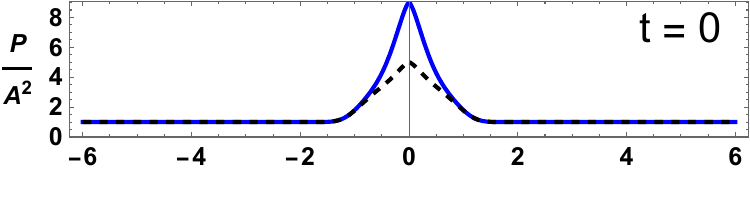}
\includegraphics[width=0.55\textwidth]{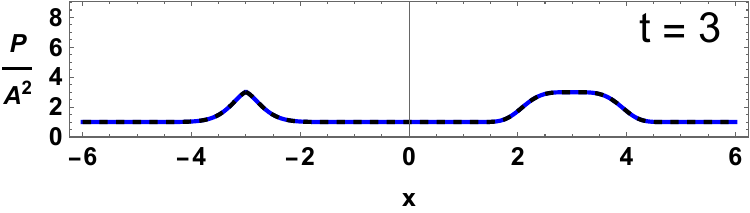}
\caption{Non-linear interaction of two solitonic waves (blue), compared to the naive prediction using linearised hydrodynamics (dashed). The background state has pressure $A^2$, the right-moving soliton has profile $f(L)=A e^{-L^4}$, and left-moving soliton has profile $g(L)=Ae^{-4L^{3/2}}$. When the solitons meet, non-linearities cause the pressure to become much larger than the sum of the pressures of the individual waves. Despite this non-linearity, the waves recover their original shape after the interaction.}
    \label{fig:Solitons}
\end{figure}

The solitonic nature of sound waves is at this point evident: Two localized wavepackets $f(x-t)$ and $g(x+t)$ can meet, overlap for a while, and then depart from each other, eventually recovering their original shape (see figure \ref{fig:Solitons}). Again, from a hydrodynamic perspective, the interaction between such wavepackets is a highly non-linear process, which is reflected in the non-linear dependence of $\{P,v\}$ on $\{f,g\}$. However, such non-linearities leave no trace in the aftermath of the interaction.

\newpage
\section{A new type of singularity}

In this section, we show that, in 3+1 dimensions, compactly-supported ultrastiff matter can evolve towards states with spacelike velocity. This is a signal that, at some point, hydrodynamics must break down. This breakdown of the fluid description is not restricted to the boundary, as there may be some instant of time where $u^\mu \propto \partial^\mu \Psi$ is spacelike everywhere in space.

\subsection{A simple theorem}

Let us study the evolution of a finite fluid undergoing free expansion in vacuum (in 3+1 dimensions). For clarity, let us assume that the initial state, at $t=0$, is static, namely $u^\mu=(1,0,0,0)$. The initial pressure profile is assumed to be a smooth compactly-supported function $P_0(\textbf{x})$. Then, we need to solve the following Cauchy problem:
\begin{equation}\label{cauchy}
\begin{cases}
\partial_\mu \partial^\mu \Psi=0 \, , \\
\Psi(0,\textbf{x})=0 \, , \\
\partial_t \Psi(0,\textbf{x})=-\sqrt{P_0(\textbf{x})} \, . \\
\end{cases}
\end{equation}
The analytical solution to the wave equation in 3+1 dimensions is well-known \cite[\S 2.4.1]{Evanbook}:
\begin{equation}
    \Psi(t,\textbf{x})= -t \int_{\mathcal{S}^2} \dfrac{d^2 \Omega}{4\pi} \sqrt{P_0(\textbf{x}{+}t\mathbf{\Omega})} \, ,
\end{equation}
where $\mathbf{\Omega}$ is a vector belonging to the unit sphere $\mathcal{S}^2$.

Now, fix a location $\textbf{x}$ inside the support of $P_0$. At $t=0$, $\Psi(0,\textbf{x})$ vanishes. Furthermore, if we pick $t_f$ larger than the initial size of the fluid, then the vector $\textbf{x}{+}t_f\mathbf{\Omega}$ falls outside the support of $P_0$, and $\Psi(t_f,\textbf{x})$ also vanishes. Thus,
\begin{equation}\label{sganazzo}
    \int_0^{t_f} \partial_t \Psi(t,\textbf{x}) dt=\Psi(t_f,\textbf{x})-\Psi(0,\textbf{x}) =0 \, .
\end{equation}
However, recalling that the Cauchy data is given by \eqref{cauchy}, and that $P_0(\textbf{x})>0$, we conclude that the derivative $\partial_t \Psi(t,\textbf{x})$ (being continuous) is negative in an open neighbourhood of $t=0$. Hence, for the integral in \eqref{sganazzo} to vanish, there must be an instant of time $t>0$ at which $\partial_t\Psi(t,\textbf{x})$ becomes positive. When this happens, $\partial^\mu \Psi$ fails to be timelike futuredirected, and the correspondence with the fluid description breaks down. This leads us to the following theorem.
\begin{theorem}\label{theo2}
If a finite ball of ultrastiff matter starts in a hydrostatic state, the vector $\partial^\mu \Psi$ exits the future lightcone in finite time at any given location.
\end{theorem}
\begin{figure}
    \centering
\includegraphics[width=0.54\textwidth]{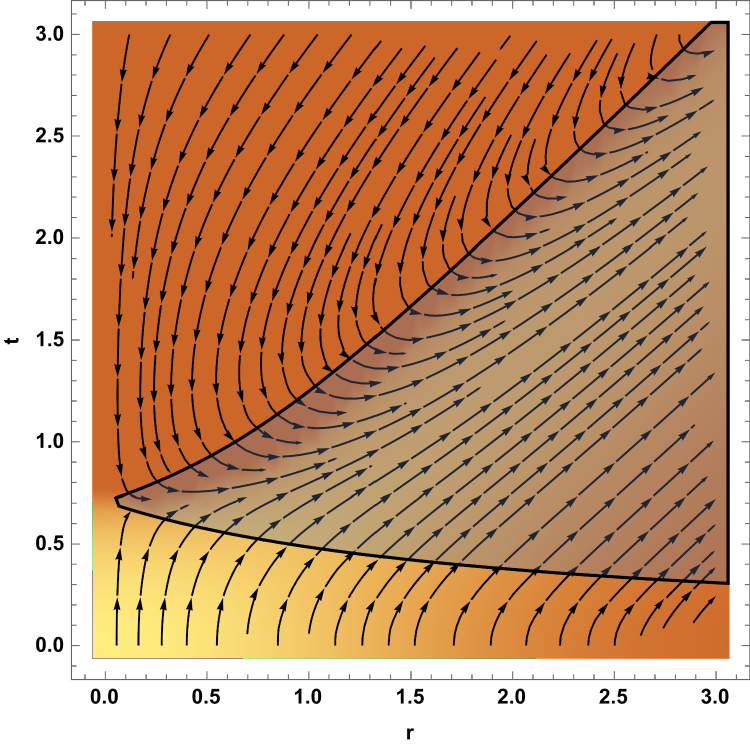}
    \caption{Minkowski diagram of a spherically symmetric freely expanding fluid, with initial pressure profile $P_0(r)=16 e^{-2r^2}$, where $r$ is the radial coordinate. The colormap represents the density of particles in the given reference frame (namely $n u^0=\partial^t \Psi$), and the black arrows mark the integral lines of $\partial^\mu \Psi$. Note that, in the upper region, where the color becomes dark orange, the density is negative (in agreement with our general argument), and, correspondingly, the worldlines point downward. The ``greyish'' domain surrounded by the continuous black line is the truly pathological region, where $u^\mu \propto \partial^\mu \Psi$ is spacelike (equivalently, $P=-\partial_\mu \Psi \partial^\mu \Psi$ is negative). 
    The scalar field dual to this process is $\Psi(t,r){=}[e^{-(r+t)^2}{-}e^{-(r-t)^2}]/r$.}
    \label{fig:negpress}
\end{figure}
An explicit example illustrating this pathological behavior is given in figure \ref{fig:negpress}.
Note that, while in the theorem $\text{``finite''}=\text{``compactly supported''}$, the main result remains true also for initial pressure profiles $P_0(\textbf{x})$ with infinite support, but whose tails decay to zero fast enough (e.g. exponentially), as is the case in the figure.

\subsection{Mechanism of singularity formation}\label{singulone}

Let us have a closer look at figure \ref{fig:negpress}. 

Initially, the fluid is at rest, i.e. $u^\mu=(1,0,0,0)$, and hydrodynamics is applicable. Then, the pressure gradients cause the fluid elements to accelerate more and more, so the flow speed approaches the speed light. Now, we are used to thinking that the actual speed-of-light limit can never be reached, as this would entail having infinite energy density. However, that is not necessarily the case. In fact, we recall that the energy density of a moving fluid is
\begin{equation}
T^{00}=\dfrac{\varepsilon+P}{1-v^2} -P\, ,
\end{equation}
so the fluid can reach  $v\rightarrow 1$ in finite time, provided that $\varepsilon+P\rightarrow 0$, which is precisely what happens in figure \ref{fig:negpress}. Along the continuous black line, we have that $\varepsilon=P \rightarrow 0$ and $u^\mu \rightarrow \infty$, so that all the components of $T^{\mu \nu}$ are ``$0 \times \infty$'' indeterminate forms which maintain a finite value. Of course, the fluid description breaks down here, since a singularity in the variable $u^\mu$ has been reached. Indeed, if we move past this line, the scalar field solution $\Psi$ enters the grey region, where $\partial^\mu \Psi$ is spacelike, so there is no way to interpret the scalar field as an ideal fluid anymore.

\subsection{Breakdown of the ultrastiff approximation}

Consider a zero-temperature fluid with equation of state $P=c_s^2 \varepsilon$, where $c_s^2$ is a constant that is very close to 1. Also in this case, there exists a notion of irrotational flow, where $\mu u^\nu = \text{const} \times \partial^\nu \Psi$. In particular, if we fix the constant factor such that $ \partial^\nu \Psi \equiv \mu u^\nu$, then the following equation of motion holds \cite[\S 3.7.4]{rezzolla_book}:
\begin{equation}\label{capiscocosasuccede}
\partial_\mu \partial^\mu \Psi +(c_s^{-2}{-}1) \partial^\mu \Psi \, \partial_\mu \ln\sqrt{- \partial_\alpha \Psi \partial^\alpha \Psi} =0 \, .
\end{equation}
Now, let us treat $\lambda=c_s^{-2}-1>0$ as a small parameter, and let us expand the solution of \eqref{capiscocosasuccede} in powers of $\lambda$, namely $\Psi=\Psi_{(0)}+\lambda \Psi_{(1)}+\mathcal{O}(\lambda^2)$. Then, we have that
\begin{equation}\label{chebuffo}
\begin{split}
\partial_\mu \partial^\mu \Psi_{(0)} ={}& 0 \, , \\
\partial_\mu \partial^\mu \Psi_{(1)} ={}& - \partial^\mu \Psi_{(0)} \, \partial_\mu \ln\sqrt{- \partial_\alpha \Psi_{(0)} \partial^\alpha \Psi_{(0)}}  \, .
\end{split}
\end{equation}
We immediately see that the expansion in $\lambda$ makes sense only if $\partial_\alpha \Psi_{(0)} \partial^\alpha \Psi_{(0)}$ stays away from zero, meaning that $\partial^\nu \Psi$ remains inside the future lightcone. In fact, if $\partial_\alpha \Psi_{(0)} \partial^\alpha \Psi_{(0)}\rightarrow 0$, the logarithm in the second line of \eqref{chebuffo} diverges, and the perturbative expansion breaks down. This suggests that the dynamics of ``real fluids'' where $P$ is \textit{almost} equal to $\varepsilon$ (like the fluids in \cite{Zeldovich:1961sbr,MooreSpeedofSound2024gmt}) can be well approximated using \eqref{idealfluid} only if we stay away from singularities like the ones mentioned in Theorem \ref{theo2}. For example, in figure \ref{fig:negpress}, we should think of the continuous black line as marking the onset of phenomena that are ``infinitely sensitive'' to the difference between $c_s^{-2}$ and 1. 

\begin{figure}[h!]
    \centering
\includegraphics[width=0.55\textwidth]{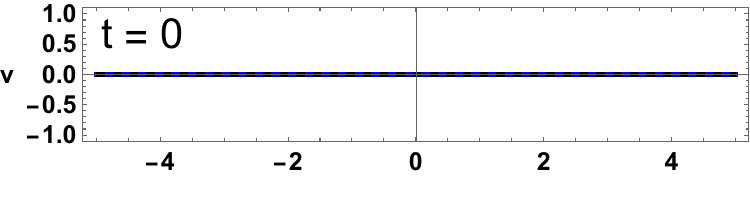}
\includegraphics[width=0.55\textwidth]{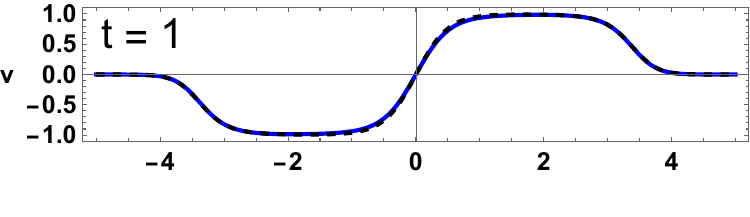}
\includegraphics[width=0.55\textwidth]{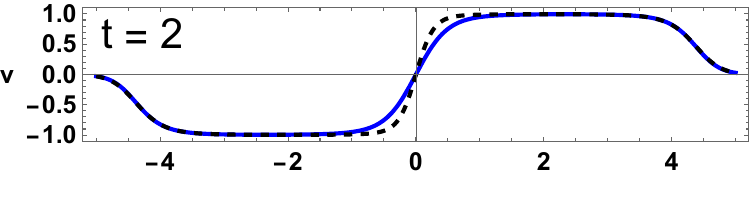}
\includegraphics[width=0.55\textwidth]{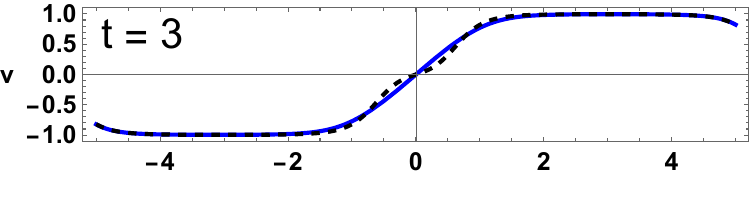}
\includegraphics[width=0.55\textwidth]{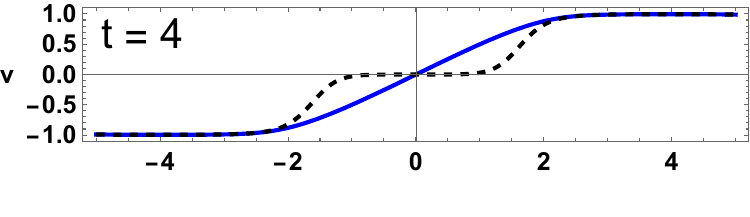}
\includegraphics[width=0.55\textwidth]{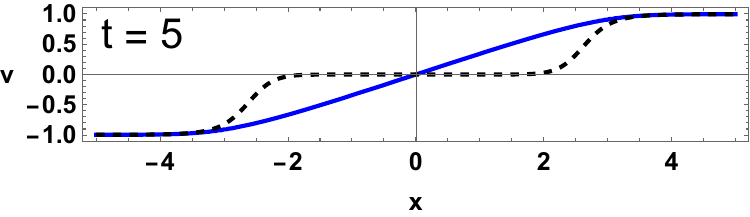}
\caption{Evolution of a fluid with initial fourvelocity $u^\mu=(1,0,0,0)$ and initial pressure $P(x)=e^{-2 x^2} (1.\, +0.004 e^{x^2})+4 \times 10^{-6}$. Each plot is a snapshot of the velocity profile $v(t,x)$ of a fluid with $c_s=0.995$ (blue line), compared with the ultrastiff solution (dashed line). Under ``normal'' conditions, the two solutions would be indistinguishable. However, the rapid expansion of the bump causes the pressure to drop very rapidly near $x=0$, and this enhances the second term in \eqref{capiscocosasuccede}. As a result, the ultrastiff model no longer describes the evolution of the fluid with $c_s=0.995$ at late times.}
    \label{fig:sing}
\end{figure}

In figure \ref{fig:sing}, we give a numerical example of a simple flow in 1+1 dimensions where $\partial_\alpha \Psi \partial^\alpha \Psi$ becomes very small in a nieghbourhood of $x=0$, which causes the full solution of \eqref{capiscocosasuccede} (with $c_s^{-2} \approx 1.01$) to considerably depart from the ultrastiff approximation\footnote{The breakdown of the perturbative solution along the flow of figure \ref{fig:negpress} implies that numerical simulations of \eqref{capiscocosasuccede} are extremely sensitive to rounding errors near the black line. This means that, if $c_s^2$ is too close to 1, numerical solutions cannot be trusted inside the grey region. Potential resolutions to this issue may exist, but investigating them falls outside the scope of the present work.}.

\section{Conclusions}
\vspace{-0.3cm}

Ultrastiff fluids showcase the ability of relativity to modify the collective dynamics of matter.
Indeed, the phenomena discussed in this article can be understood from basic special-relativity principles via the following thought experiment. Consider a 1+1 dimensional configuration, where a small sound wavepacket with position $x(t)$ travels on top of a non-linear background flow $v(t,x)$. In the eikonal approximation \cite[\S 53]{landau2}, we have the equation of motion 
\begin{equation}\label{eikonal}
\dfrac{dx}{dt}= \dfrac{c_s+v}{1+c_s v} \, .
\end{equation}
We see that, if $c_s=1$, the background flow does not affect the motion of the wavepacket, and we have that $dx/dt=1$. This intuitively explains why non-linear sound waves behave like solitons in ultrastiff matter (see figure \ref{fig:Solitons}): The non-linear interaction between two waves cannot cause one incoming wave to ``disperse'', because every part of such wave moves at the same speed (namely 1) also in the presence of the other incoming wave.

The above discussion also clarifies why a fluid with speed of sound very close to 1 is fundamentally different from a fluid with $c_s^2 \, {\equiv}\, 1$ (see figure \ref{fig:sing}). In fact, if $c_s^2\,{<}\, 1$, and the background flow is sufficiently fast (specifically, $v{\leq} {-}c_s$), it is always possible to reverse the direction of sound-wave propagation in the lab frame (i.e. $dx/dt\leq 0$). This, on the other hand, is not possible when $c_s^2$ is exactly 1, since in that case $dx/dt =1$ for any $v$. This kind of ``high sensitivity'' of hydrodynamics to the difference $c_s^{-2}-1$ becomes especially significant in regions where the pressure $P$ tends to zero. In fact, only in such cases the velocity $v$ can approach $\pm 1$ while maintaining a finite $T^{00}$ (see section \ref{singulone}).

In summary, the study of ultrastiff fluids offers a unique window into regimes where relativistic effects dominate fluid dynamics in unexpected ways. Future work could investigate how the (effectively linear) phenomena described in section \ref{IIIyup} change in systems where $c_s^2$ approaches but does not exactly reach the speed of light. In particular, it would be interesting to examine how small viscous or quantum corrections modify the soliton picture, and whether similar mechanisms operate in astrophysical contexts like neutron stars or in early-universe cosmology.

\section*{Acknowledgements}

This work is partially supported by a Vanderbilt's Seeding Success Grant. I thank M. Disconzi, J. Speck, L. Abbrescia, and 
H. Freist\"{u}hler for reading the manuscript and providing useful feedback.

\appendix

\bibliography{Biblio}

\label{lastpage}

\end{document}